\documentclass[12pt]{article}
\usepackage{times}
\usepackage{graphicx}
\usepackage{proteins,citesupernumber}

\begin{document}
\setlength{\baselineskip}{20pt}
\begin{flushleft}
{\Large \bf Predicting Residue-wise Contact Orders of Native Protein Structure from  Amino Acid Sequence}

\vspace{5mm}
Akira R. Kinjo$^{1,2,*}$ and Ken Nishikawa$^{1,2}$

\vspace{3mm}
$^{1}$Center for Information Biology and DNA Data Bank of Japan,
National Institute of Genetics, Mishima, 411-8450, Japan\\
$^{2}$Department of Genetics, 
The Graduate University for Advanced Studies (SOKENDAI), 
Mishima, 411-8540, Japan

\vspace{1cm}
$^{*}$Correspondence to Akira R. Kinjo.\\
Center for Information Biology and DNA Data Bank of Japan, National Institute of Genetics, Mishima, 411-8540, Japan\\
Tel: +81-55-981-6859, Fax: +81-55-981-6889\\
E-mail: akinjo@genes.nig.ac.jp

\vspace{1cm}
Running title: Residue-wise contact order prediction.

\vspace{1cm}
\end{flushleft}
\newpage
\begin{abstract}
Residue-wise contact order (RWCO) is a new kind of one-dimensional protein 
structures which represents the extent of long-range contacts. 
We have recently shown that a set of three types of one-dimensional structures 
(secondary structure, contact number, and RWCO) contains sufficient 
information for reconstructing the three-dimensional structure of proteins.
Currently, there exist prediction methods for secondary structure and contact 
number from amino acid sequence, but none exists for RWCO. Also, the properties of 
amino acids that affect RWCO is not clearly understood. Here, we present a
linear regression-based method to predict RWCO from amino acid sequence, 
and analyze the regression parameters to identify the properties that 
correlates with the RWCO. The present method achieves the significant 
correlation of 0.59 between the native and predicted RWCOs on average. 
An unusual feature of the RWCO prediction is the remarkably large optimal 
half window size of 26 residues.
The regression parameters for the central and near-central residues of the 
local sequence segment highly correlate with those of the contact 
number prediction, and hence with hydrophobicity.
\end{abstract}
\newpage
\section*{Introduction}
One of the main goals of protein structure prediction is to provide an 
intuitive picture of the relationship between the amino acid sequence
and the native three-dimensional (3D) structure of proteins.
To this end, a number of methods have been developed for \textit{ab initio} or 
\textit{de novo} protein structure prediction. However, such methods are 
usually very complicated and make it difficult to intuitively understand 
the relationship between amino acid sequence and 3D structure.
In this respect, one-dimensional (1D) structures\cite{Rost2003} of 
proteins may be conventional intermediate representations of both 
sequence and structure 
of proteins as it is easy to grasp the correspondence between sequence and 
structural characteristics. 

Since 1D structures are 3D structural features projected onto strings of 
residue-wise structural assignments\cite{Rost2003}, a large part of 3D 
information appears to be lost. That is, the correspondence between amino 
acid sequence and 1D structures does not seem to be sufficient for 
uncovering the correspondence between amino acid sequence and 3D structure.
However, Porto \textit{et al.}\cite{PortoETAL2004} have recently shown that 
the contact matrix of a protein structure can be uniquely recovered from 
its principal eigenvector. Since the protein 3D structure can be recovered 
from the contact matrix\cite{VendruscoloETAL1997}, the result of 
Porto \textit{et al.}\cite{PortoETAL2004} indicates that 
the information contained in the 3D structure can be expressed as a 
one-dimensional representation.
Furthermore, we have recently shown that 3D structure of proteins can be 
reconstructed from a set of three types of 
1D structures\cite{KinjoANDNishikawa2005}. 
In other words, the 3D structure of a protein is essentially equivalent 
to a set of three types of 1D structures. 
These 1D structures are namely secondary structure, contact number and 
residue-wise contact order.
The fact that the 3D structure of a protein can be recovered from 
a set of these 1D structures
opens a new possibility for elucidating the sequence-structure relationship
of proteins.

The secondary structure of a protein is a string of symbols representing 
$\alpha$ helix, $\beta$ strand, or coils. The contact number of each residue 
in a protein is defined by the number of contacts the residue makes with other 
residues in the protein. More precisely, 
if we represent the contact map of the protein by $C_{i,j}$ ($C_{i,j} = 1$ 
if the $i$-th and $j$-th residues are in 
contact, or $C_{i,j} = 0$ otherwise), the contact number $n_{i}$ 
of the $i$-th residue is defined by $n_{i} = \sum_{j}C_{i,j}$. 
Similarly, the residue-wise contact order (RWCO) $o_{i}$ 
of the $i$-th residue of a protein is defined 
by $o_{i} = \sum_{j}|i-j|C_{i,j}$,
that is, a sum of sequence separations between the residue and the 
contacting residues\cite{KinjoANDNishikawa2005}.
The contact order was first introduced as a per-protein quantity by 
Plaxco et al.\cite{PlaxcoETAL1998} to study the correlation between protein 
topology and folding rate. The RWCO introduced here is a generalization of
 the contact order, and is a per-residue quantity. 

At least in principle, if we can predict those 1D structures, 
we can also construct the corresponding 3D structures. Many accurate methods
have been developed for secondary structure prediction\cite{Rost2003}. 
We have developed a method to predict the contact number from amino acid 
sequence\cite{KinjoETAL2005} with the average correlation of 0.63 
between the native and predicted contact numbers. However, there is 
no method for predicting RWCO from amino acid sequence to date, 
and it is not clear if the prediction is possible at all. 
The primary objective of the present paper is to develop a method 
to predict RWCO from amino acid sequence.

While the accurate prediction of structural properties is important for 
its own sake, for a thorough understanding of the 
sequence-structure relationship, we still need to identify the properties 
of amino acid sequence that determine the structure.
From the vast amount of studies on secondary structure prediction in the past, 
we are now convinced 
that each amino acid has a particular propensity for a particular secondary 
structure, although the final secondary structures in the native 
structure are determined in the global context. Also, contact number is 
 closely related to the hydrophobicity of amino acids. Thus, both 
secondary structure and contact number have clear connections with
the properties of amino acids. As for the residue-wise contact order, 
 its geometrical meaning is clear (i.e., a quantity related to the extent of 
long-range contacts), but the conjugate properties of amino acids are not.
As the second objective of the present study, we attempt to identify the 
amino acids' property affecting RWCO by examining the parameters derived 
for the prediction method.

The prediction method developed in this paper is based on a simple linear 
regression scheme which was also applied to the contact number 
prediction in our previous study\cite{KinjoETAL2005}. 
By examining the regression parameters, 
we show that the RWCO is primarily determined by the pattern of hydrophobicity
of amino acids. 
Although the method is extremely simple, it yields a significant 
correlation of 0.59 between the native and predicted RWCOs. 
While further refinement is definitely necessary to apply the method 
for 3D structure prediction, the present method will serve as a basis for 
more elaborate methods yet to be developed.

\section*{Materials and Method}
\subsection*{Definition of residue-wise contact order}
As mentioned in the Introduction, the residue-wise contact order (RWCO) of 
the $i$-th residue is defined by
\begin{equation}
o_{i} = \frac{1}{L}\sum_{j:|j-i|>2}|i-j|C_{i,j}\label{eq:def}
\end{equation}
where the summation is normalized by the length $L$ of the amino acid 
sequence of the protein and $C_{i,j}$ represents the contact map of the protein.
We exclude trivial contacts between nearest- and next-nearest residues 
along the sequence.
To make the RWCO useful for molecular dynamics simulations, the contact 
between two residues is defined by a smooth sigmoid function:
\begin{equation}
C_{i,j} = 1/\{1+\exp[w(r_{i,j} - d_c)]\}
\end{equation}
where $r_{i,j}$ is the distance between $C_{\beta}$ atoms of the $i$-th 
and $j$-th 
residues ($C_{\alpha}$ atoms for glycine), $d_c$ is the cut-off distance for 
the contact definition, and $w$ is 
a parameter that determines the sharpness of the sigmoid function. 
To be consistent with our previous 
studies\cite{KinjoETAL2005,KinjoANDNishikawa2005},
we set $d_c = 12$\AA{} and $w=3$ throughout the present paper.

We also define the normalized (relative) RWCO by
\begin{equation}
{y}_{i}^{p} = ({o}_{i}^{p} - \langle {o}_{i}^{p} \rangle)/
\sqrt{\langle({o}_{i}^{p} - \langle {o}_{i}^{p} \rangle)^2\rangle}
\label{eq:normal}
\end{equation}
where $\langle \cdot \rangle$ denotes averaging operation over the given 
protein chain $p$.

\subsection*{Prediction scheme}
To predict the RWCO of each residue in a protein, we first conduct three 
iterations of PSI-BLAST\cite{AltschulETAL1997} search against the 
NCBI non-redundant amino acid sequence database to obtain the sequence profile 
of the protein with the E-value cut-off of $10^{-7}$. 
We use the amino acid score table of the  
PSI-BLAST  profile which is represented as $f(i,a)$ 
($i$: site, $a$: amino acid) in the following (instead of the frequency table 
used in the previous study\cite{KinjoETAL2005}).

The RWCO $\hat{o}_{i}^{p}$ of the $i$-th residue in the protein $p$ 
is predicted in two steps. First we predict the normalized RWCO $y_{i}^p$ for 
each residue, and then we combine it with the mean $\mu^p$ and standard 
deviation (S.D.) $\sigma^p$ of the RWCOs of the protein, 
which are predicted separately. The normalized RWCO is predicted by 
the following linear regression scheme:
\begin{equation}
\hat{y}_{i}^{p} = \sum_{m=-M}^{M}\sum_{a}^{\mbox{\scriptsize residue types}}C_{m,a}f^{p}(i+m,a) + C\label{eq:reg}
\end{equation}
where $M$ is the half window size (a free parameter to be determined), 
$f^{p} (i+m,a)$ represents an element of the PSI-BLAST profile of the 
protein $p$, and $C_{m,a}$ and $C$ are regression parameters.
Both amino and carboxyl termini are treated by introducing an extra symbol 
for the ``terminal residue.'' 
Thus, the RWCO of the $i$-th residue is expressed as a linear function of 
the local sequence of $2M+1$ residues surrounding the $i$-th residue.

The values of $C_{m,a}$ and $C$ are determined so as to minimize the prediction
error over a database of protein structures. The error function is defined by
\begin{equation}
E = \sum_{p}\sum_{i}(y_{i}^{p} - \hat{y}_{i}^{p})^{2}
\end{equation}
where $y_{i}^{p}$ is the observed normalized RWCO of the $i$-th residue of the 
protein $p$.
The minimization of $E$ can be achieved by the usual least squares method.

The mean ($\mu^p$) and standard deviation ($\sigma^p$) of 
the RWCOs of a protein are predicted from the amino acid 
composition ($f_a^p$) and sequence length ($L^p$) of the protein $p$ in 
the same manner as we have done for the contact number 
prediction\cite{KinjoETAL2005}.
That is, the mean and S.D. are predicted by the following linear regression 
scheme:
\begin{eqnarray}
  \hat{\mu}^{p} & = &\sum_{a}A_{a}f_{a}^{p} + A_{l}F(L^{p}) + A\\
  \hat{\sigma}^{p} & = &\sum_{a}D_{a}f_{a}^{p} + D_{l}F(L^{p}) + D
\end{eqnarray}
where $F(L^p) = L^p$ for $L^p < 300$ and $F(L^p) = 300$ for $L^p \geq 300$,
and $A_{a}, A, D_{a}, D$ are regression parameters.
The final value for the predicted absolute RWCO ($\hat{o}_{i}^{p}$) is given by 
\begin{equation}
  \hat{o}_{i}^{p} = \hat{\mu}^{p} + \hat{\sigma}^{p}\hat{y}_{i}^{p}.
\label{eq:pred}
\end{equation}

\subsection*{Data set}
We first selected representative proteins from each superfamily of 
all-$\alpha$, all-$\beta$, $\alpha/ \beta$, $\alpha + \beta$, 
and multi-domain classes of the SCOP\cite{SCOP} (version 1.65) protein 
structure classification database through the ASTRAL\cite{ASTRAL}  
database. Those structures which were present in this superfamily 
representative set but were absent from the 40\% representative set of 
ASTRAL, those containing chain breaks (except for termini), or those 
with the average contact number of less than 7.5 (non-compact structures) 
were discarded. 
Non-standard amino acid residues were converted to the corresponding standard
residues when possible, otherwise discarded. 
When $C_\beta$ atoms were absent in non-glycine residues, they were modeled 
by the SCWRL\cite{SCWRL3} side-chain prediction program.
After all, there remained 680 protein chains. The list of this data set 
will be available from the author's website.

For training the parameters and testing the prediction accuracy, we performed 
a 15-fold cross-validation test. The 680 proteins were randomly 
divided into two groups, one consisting of 630 proteins for training 
the parameters (training set), and the other (test set) consisting of 
50 proteins for testing the prediction using the parameters obtained from 
the training set. The procedure was iterated for 15 times.

\subsection*{Measures of prediction accuracy}
We employ two measures for evaluating the prediction accuracy. 
The first one is the correlation coefficient ($Cor_p$) between the observed 
and predicted RWCOs for a given protein $p$, which is defined by
\begin{equation}
  Cor_{p} = \frac{\langle (o_{i}^{p} - \langle o_{i}^{p} \rangle)(\hat{o}_{i}^{p} - \langle \hat{o}_{i}^{p}\rangle)\rangle}{
\sqrt{\langle (o_{i}^{p} - \langle o_{i}^{p} \rangle)^{2}\rangle} 
\sqrt{\langle (\hat{o}_{i}^{p} - \langle \hat{o}_{i}^{p} \rangle)^{2}\rangle}}.
\label{eq:cor}
\end{equation}
The $Cor_p$ measures the consistency of the normalized RWCOs.
In order to measure the accuracy of the predicted absolute values, we 
use the RMS error divided by the standard deviation of the observed
RWCO ($DevA_p$):
\begin{equation}
    DevA_{p} = \frac{\sqrt{\langle (o_{i}^{p} - \hat{o}_{i}^{p})^{2}\rangle}}
{\sqrt{\langle (o_{i}^{p} - \langle o_{i}^{p} \rangle)^2\rangle}}.
\label{eq:deva}
\end{equation}

\section*{Results}
\subsection*{Optimal window size}
\begin{figure}
  \begin{center}
\includegraphics[width=8cm]{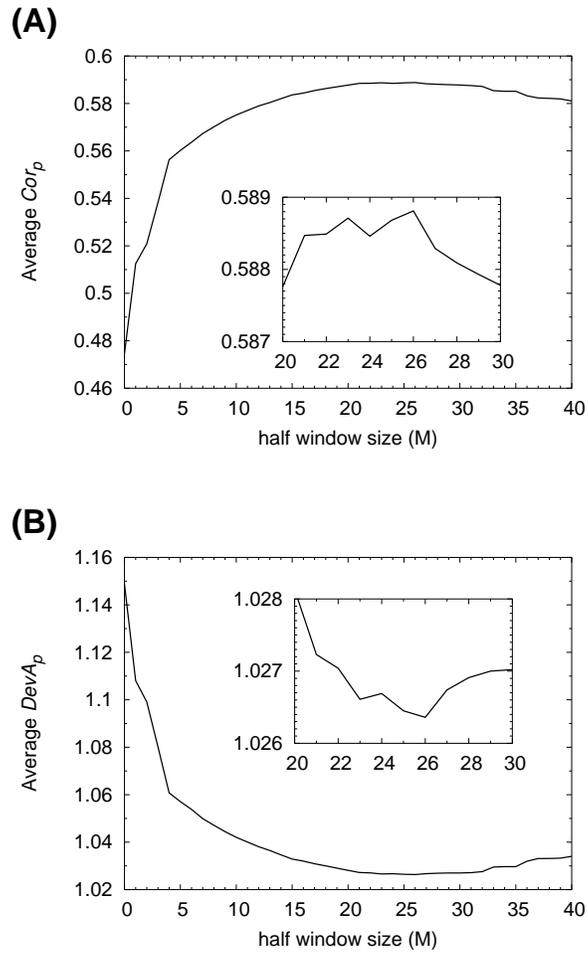}
  \end{center}
\caption{\label{fig:window}Prediction accuracy as a function of window size.
(A) The correlation coefficient ($Cor_p$) between the native and predicted RWCO, averaged
over the test set proteins. (B) Deviation of the predicted RWCO from the native one ($DevA_p$), averaged over the test set proteins.}
\end{figure}

In the prediction scheme presented in this paper, the half window size $M$ 
is a free parameter. We determine its value so that the prediction accuracy 
is maximized. We have performed a 15-fold cross-validation test with $M$ 
ranging from 0 to 40. The result is summarized in Figure~\ref{fig:window}. 
The correlation coefficient $Cor_p$ (averaged over the test sets) 
ranges from 0.48 at $M=0$ to $\approx$ 0.59 at $M=26$ 
(Figure \ref{fig:window} A). It should be noted that the correlation of 0.48 
is already statistically significant given the 
average sequence length (172 residues) of the proteins in the data set.
The value of  $Cor_p$ monotonically increases from $M=0$ to $M=26$, but 
starts to saturate for $M > 20$ and decreases slowly for $M>26$. 
The deviation $DevA_p$ 
(averaged over the test sets) shows a consistent trend with $Cor_p$ 
(Figure \ref{fig:window} B), and it reaches the minimum value of $\approx$ 
1.03 at $M=26$.
Thus, the optimal window size has been determined to be $M=26$.

This optimal window size of $M=26$ is much larger than the ones for any 
other 1D structure predictions. As far as we are aware, this is the longest 
range of correlation observed between 1D structure and amino acid sequence.
For example, the optimal half window size  is 
$M=9$ for contact number prediction (see below) and $M = 6-8$ for secondary 
structure prediction. Large window sizes usually result in over-fitting  
the training data, but such is not the case for RWCO prediction, as we have 
performed cross-validation tests. This unusually long-range correlation with 
amino acid sequence is a conspicuous property of the RWCO. 

\subsection*{Distribution of correlation}
\begin{figure}
  \includegraphics[width=8cm]{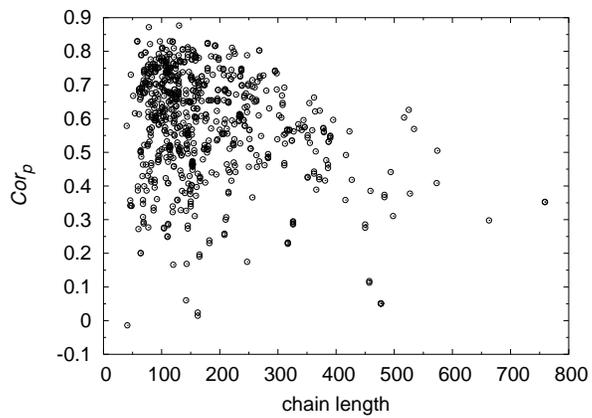}
\caption{\label{fig:len_cor}$Cor_p$ plotted against chain length. Each point represents a protein in one of the test sets.}
\end{figure}

As indicated by the average values of $Cor_p$ and $DevA_p$, 
the linear regression method with $M=26$ tends to produce more accurate
predictions than with other window sizes. However, the prediction 
accuracies for individual proteins do differ significantly as shown in 
Figure \ref{fig:len_cor}. While most of the proteins
are decently predicted with correlations of 0.5 or higher, 
some proteins exhibit very poor correlations. The poorly predicted proteins 
are found not well-packed due to the small size of the protein (e.g., 
SCOP domain d1fs1a1), 
 a large fraction of structurally disordered regions (e.g., d1cpo\_1), or 
being a subunit of a large complex (e.g., d1mtyg\_). 

The prediction accuracy does not strikingly differ depending on the structural
class of proteins (Table \ref{tab:histo}). However, all-$\alpha$ proteins 
show slightly poorer correlations compared to other classes, 
and $\alpha + \beta$ proteins show relatively better correlations. The latter
may be due to the over-dominance of the $\alpha + \beta$ proteins in the 
data sets.

\begin{figure}
  \begin{center}
    \includegraphics[width=7cm]{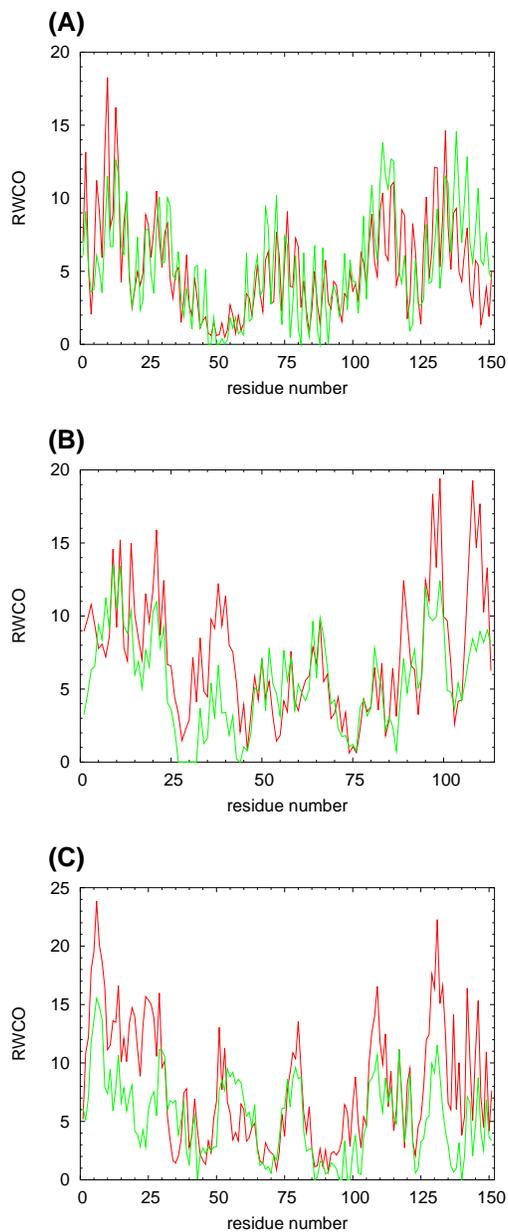}
  \end{center}
\caption{\label{fig:ex}Examples of prediction. Red: native RWCO; Green: predicted RWCO.
(A) SCOP domain d1a6m\_\_ (myoglobin, all-$\alpha$), $Cor_p = 0.73$, 
$DevA_p = 0.75$; 
(B) SCOP domain d1ifra\_ (Lamin A/C globular tail domain, all-$\beta$), 
$Cor_p =0.72$, $DevA_p = 0.87$;
(C) SCOP domain d1a9xb1 (Carbamoyl phosphate synthetase, small subunit N-terminal domain, $\alpha / \beta$), $Cor_p = 0.72$, $DevA_p = 0.81$. }
\end{figure}

In Figure \ref{fig:ex}, three examples of predicted RWCO are shown. 
Despite the relatively good correlation between the native and predicted RWCOs,
the absolute values of predicted RWCOs at many sites 
significantly differ from the corresponding native RWCOs. 
This behavior is indicated by the relatively
large value of $DevA_p \approx 1.03$ (Figure \ref{fig:window} B). 
In particular, we notice that  RWCOs of large values are consistently  
underestimated. This behavior suggests that some cooperative effects 
be taken into account for better prediction. 
Provided that the present method is based exclusively
on one-body terms (Eq. \ref{eq:reg}), the prediction accuracy achieved is
satisfactory, at least qualitatively. 

\subsection*{Regression parameters as functions of sequence position}
\begin{figure}
  \begin{center}
\includegraphics[width=16cm]{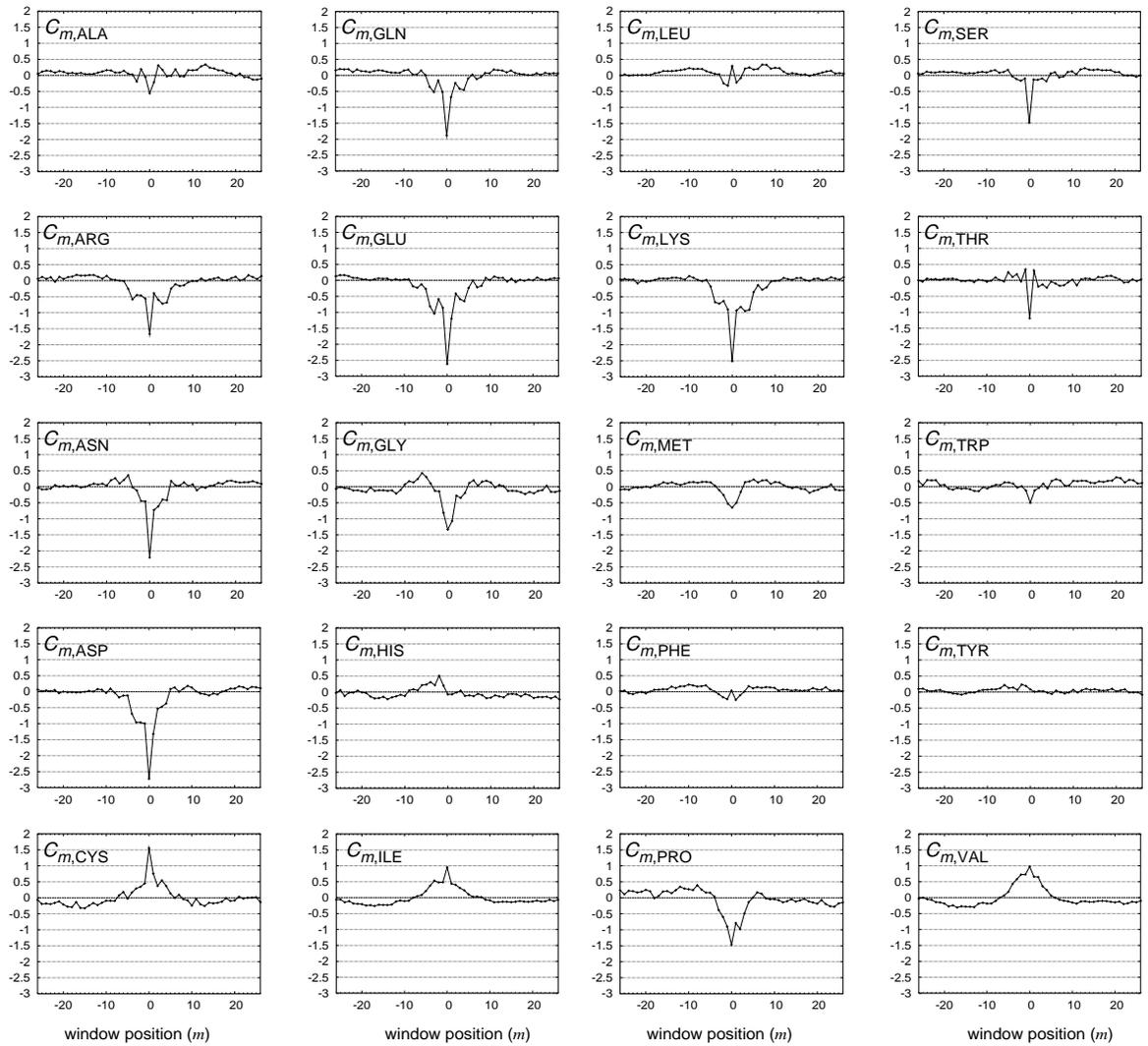}    
  \end{center}
\caption{\label{fig:aaprop}$C_{m,a}$ for each amino acid type ($a$) as a function of the window position ($m$).}
\end{figure}

Since the present study is the first attempt to develop a prediction method 
for RWCO, it is of interest to examine the properties of amino acid
residues that affect the RWCO, which are reflected in the values of 
the regression coefficients $C_{m,a}$. 
Figure \ref{fig:aaprop} shows the values of $C_{m,a}$ for each amino acid 
type $a$ as a function of the window position $m$. For all the amino acid 
types, the peak of $C_{m,a}$, when present, is at the center ($m=0$). 
We can easily recognize that these values, those at $m=0$ in particular, 
are related to the hydrophobicity of amino acids. 
That is, $C_{0,a}>0$ for hydrophobic residues and $C_{0,a}<0$ for hydrophilic 
residues. 
When the amino acid index (AAindex) database\cite{TomiiANDKanehisa1996} 
was scanned for indices that 
highly correlates with $C_{0,a}$, we have found various hydrophobicity scales 
with correlations with $C_{0,a}$ over 0.90 (data not shown).  
Therefore, we can conclude that the RWCO is primarily determined by the 
pattern of hydrophobicity along the sequence.

Some amino acid types exhibit oscillation with the periodicity of 
3 to 4 residues, which is expected for the $\alpha$ helix. In fact, 
such residues (e.g., GLU, GLN, ALA, etc.) are of high $\alpha$ helix 
propensity. On the contrary, the residues of high $\beta$ strand 
propensity (e.g., ILE, VAL, etc.) do not exhibit such oscillation. 
Therefore, in addition to the hydrophobic properties, the parameters for 
RWCO also contain information for secondary structures.

\section*{Discussion}
\subsection*{Comparison with contact number prediction}
As can be seen from their definitions, the native RWCOs and contact numbers 
show a high correlation of 0.7 
(data not shown). This is also consistent with the finding that RWCOs are 
primarily determined by hydrophobicity. Because of the correlation 
between RWCO and contact number, it is of interest to ask whether it is possible to ``predict''
RWCOs using contact number prediction, and vice versa.
The result of this ``cross-prediction'' is listed in Table \ref{tab:cnrwco}.
Here, the contact number prediction\cite{KinjoETAL2005} is based 
on exactly the same linear regression scheme as the RWCO prediction method.
In order to make consistent the quality of the two different prediction 
methods, we have determined the regression parameters and the optimal 
half window size for the contact number prediction using the same training 
and test data sets as used here. 
The resulting contact number prediction method yields the average 
prediction accuracy of $Cor_p \approx 0.70$ and 
$DevA_p \approx 0.803$ with the optimal half window size of 9 
(Table \ref{tab:cnrwco}, Case B), a remarkable improvement over our 
previous study 
($Cor_p \approx 0.63$ and $DevA_p \approx 0.941$)\cite{KinjoETAL2005} 
which is likely to be due to the use of PSI-BLAST score profiles 
(we used frequency profiles derived from the HSSP database\cite{HSSP}
in the previous study). 
When the values obtained from the contact number prediction are compared 
to the native RWCOs, the highest correlation is 0.50 with the optimal half 
window size 
of $M = 4$ (Table \ref{tab:cnrwco}, Case C). Although the correlation of 0.50
is statistically significant, 
the value is much lower than the one 
obtained for the proper prediction of RWCO, $Cor_p \approx 0.59$
(Table \ref{tab:cnrwco}, Case A). For the ``prediction'' in the opposite 
direction, that is, when the values obtained from the RWCO prediction are 
compared to the native contact numbers, the correlation is as high as 0.62
with the optimal half window size of $M=4$ (Table \ref{tab:cnrwco}, Case D). 
Again, this value, though statistically significant, is lower than the 
proper contact number prediction ($Cor_p \approx 0.70$).
Interestingly, for the Cases C and D in 
Table \ref{tab:cnrwco}, the optimal half window sizes coincide ($M = 4$). 
Therefore, it is expected that the contact number and RWCO are very closely 
related with each other in terms of the short-range pattern of the 
local amino acid sequence. In other words, the distinction between the 
contact number and RWCO originates from the interactions of longer range.

\begin{figure}
  \includegraphics[width=8cm]{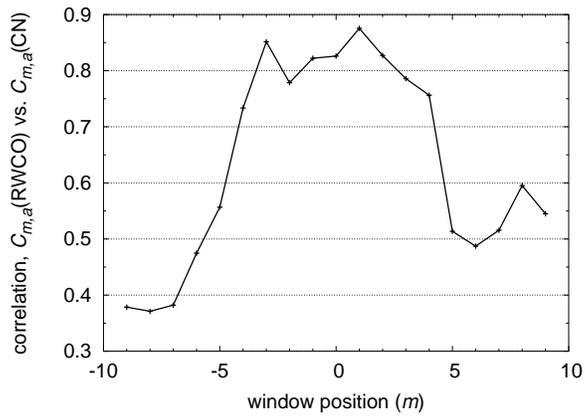}
\caption{\label{fig:parcor}Correlation between the regression parameters 
$C_{m,a}$ for contact number and RWCO predictions for each window position.
The horizontal axis is the window position $m$ in the local sequence. 
The vertical axis is the correlation coefficient between the regression 
parameters $C_{m,a}$ for RWCO prediction and those for contact number 
prediction at the window position $m$.}
\end{figure}

To further clarify the correlation between 
 RWCO and contact number predictions, we compared the regression 
parameters $C_{m,a}$ for RWCO and contact number predictions up to the 
half window size of $M=9$ (Figure \ref{fig:parcor}). 
It can be clearly seen that the both sets of regression parameters
very significantly correlate (correlation of $>0.7$) 
with each other within the window positions of 
$-4 \leq m \leq 4$ (Figure \ref{fig:parcor}), which confirms 
the above observation (Table \ref{tab:cnrwco}, Cases C and D).

\subsection*{Perspective for improving prediction accuracy}
The method for predicting RWCOs from amino acid sequence  
developed in this paper is a very primitive one.
While the correlation of 0.59 between the native and predicted 
RWCOs is significant, it is not as high as 0.70 in the case of the 
contact number prediction (Table \ref{tab:cnrwco}) 
based on the same linear regression scheme.
Furthermore, the agreement of absolute RWCO values 
is relatively poor, especially so for RWCOs of large values. 
As mentioned above, inclusion of many-body
effects seems mandatory for better RWCO prediction. 
A popular method for dealing with many-body terms is artificial 
neural networks. Other non-linear regression schemes such as radial basis
or support vector regressions can be also 
applicable.
Neural network methods as well as a support vector regression method
have been successfully applied to real value prediction of solvent
accessibility\cite{AhmadETAL2003,AdamczakETAL2004,YuanANDHuang2004}. 
Solvent accessibility is closely related to the hydrophobicity of amino 
acids, and hence is likely to be related to the RWCO. Thus, we can expect 
such non-linear regression approaches may be also useful for predicting RWCO.
However, since the RWCO prediction requires rather long segment of local 
amino acid sequence (half window size of $M=26$), 
straightforward application of non-linear regression methods requiring 
a great number of parameters may not work.
The number of parameters must be somehow reduced.
How to extract essential parameters for RWCO prediction is left for 
future studies.

An alternative route to the improved accuracy is to properly treat 
the large deviation of RWCOs along the amino acid sequence. 
For the contact number, its average over a local segment 
tends to be close to the average over the whole sequence, 
whereas, for the RWCO, such is not the case. 
For example, for the SCOP domain d1a9xb1 (Figure \ref{fig:ex}C), 
the average contact number for the whole domain, for residues 1 to 20, 
and for residues 51 to 70 are, respectively, 25.5, 28.4, and 26.6, whereas 
the corresponding averages of the RWCOs are 8.0, 14.3, and 4.9, respectively.
Since the present method is based on the globally normalized RWCO
(Eq. \ref{eq:normal}), such large deviations are difficult to 
handle. If this limitation is overcome, better prediction accuracy may be 
obtained.

\section*{Acknowledgment}
The authors thank Satoshi Fukuchi, Yoshiaki Minezaki, and Yasuo Shirakihara
for helpful comments.
Most of the computations were carried out at the supercomputing facility of
National Institute of Genetics, Japan. This work was supported in part by a
grant-in-aid from the MEXT, Japan.

The list of the SCOP domain identifiers used in the present study, and
the optimal parameter sets are available at the URL
http://maccl01.genes.nig.ac.jp/\~{}akinjo/rwco/.


\newpage
\begin{table}
\caption{\label{tab:histo}Distribution of $Cor_p$ for each SCOP class$^a$.}
\begin{center}
  \begin{tabular}{lrrrrr}\hline
range$^b$ &\multicolumn{5}{c}{SCOP class$^c$}\\
($Cor_p$) & a & b & c & d & e\\\hline
(-1,0.2]  &  4(3)  &    1(0.6)&    7(4) &    2(0.8) &    0 \\
(0.2,0.4] & 23(14) &   17(10) &   14(8) &   22(9) &    1(5) \\
(0.4,0.6] & 61(38) &   54(33) &   55(33) &   72(30) &   11(61) \\
(0.6,0.8] & 73(45) &   86(52) &   82(49) &  136(57) &    6(33) \\
(0.8,1.0] &  1(0.6)&    6(4)  &    8(5) &    8(3) &    0 \\
total     &  162 &  164 &  166 &  240 &   18\\
\hline
  \end{tabular}
\end{center}
$^a$ The number (percentage in the parentheses) 
of occurrences of $Cor_p$ for the proteins in the test sets, 
classified according to the SCOP database.\\
$^b$ The range ``$(x,y]$'' denotes $x < Cor_p \leq y$.\\
$^c$ a: all-$\alpha$, b: all-$\beta$, c: $\alpha / \beta$, d: $\alpha + \beta$, 
e: multi-domain.
\end{table}
~\\
\newpage
\begin{table}
  \caption{\label{tab:cnrwco}Cross-prediction between residue-wise contact orders and contact numbers.}
  \begin{center}
    \begin{tabular}{cccrrr}\hline
Case & Train$^a$ & Test$^b$ & $M^c$ & $Cor_p$ & $DevA_p$  \\\hline
A    &  RWCO & RWCO & 26  & 0.59 & 1.03 \\
B    &  CN   & CN   & 9  & 0.70 &  0.803\\
C    &  CN   & RWCO & 4  & 0.50 & N.A.$^d$  \\
D    &  RWCO & CN   & 4  & 0.62 & N.A.$^d$  \\\hline
    \end{tabular}
  \end{center}
$^a$Target values for which the regression parameters were trained. ``RWCO'' and ``CN'' 
indicate that the regression parameters were trained to fit the residue-wise contact orders 
and contact numbers, respectively.\\
$^b$Target values for which the ``prediction'' was applied. ``RWCO'' and ``CN'' indicate 
that predicted values were compared with the native residue-wise contact orders and native 
contact numbers, respectively.\\
$^c$Optimal half window size for the prediction.\\
$^d$Not applicable because the ranges of RWCO and CN values are different.
\end{table}
~\\

~\\
\newpage

~\\

~\\
\newpage
\end{document}